\documentclass[pdflatex,sn-mathphys,referee,a4paper]{sn-jnl}
\usepackage{siunitx}
\usepackage{physics}
\usepackage{graphicx,hyperref,amsmath,url}
\usepackage{amssymb,amsmath}
\usepackage[capitalise]{cleveref}
\DeclareSIUnit\angstrom{\text {Å}}

\jyear{2021}
\unnumbered

\newcommand{\VLD}{V_\mathrm{LD}}
\newcommand{\VRD}{V_\mathrm{RD}}
\newcommand{\VL}{V_\mathrm{L}}
\newcommand{\VR}{V_\mathrm{R}}
\newcommand{\IL}{I_\mathrm{L}}
\newcommand{\IR}{I_\mathrm{R}}
\newcommand{\GRR}{G_\mathrm{RR}}
\newcommand{\GLL}{G_\mathrm{LL}}
\newcommand{\GRL}{G_\mathrm{RL}}
\newcommand{\GLR}{G_\mathrm{LR}}
\newcommand{\VPG}{V_\mathrm{PG}}
\newcommand{\muL}{\mu_\mathrm{LD}}
\newcommand{\muR}{\mu_\mathrm{RD}}
\newcommand{\nL}{n_\mathrm{LD}}
\newcommand{\nR}{n_\mathrm{RD}}

\newcommand{\nn}{\nonumber \\} 
\newcommand{\tp}{ ^{\intercal} }
\newcommand{\dg}{^{\dagger}}

\newcommand{\half}{\frac{1}{2}}

\begin{document}

\title{Realization of a minimal Kitaev chain in coupled quantum dots}

\author*[1]{\fnm{Tom}~\sur{Dvir}}\email{tom.dvir@gmail.com}
\equalcont{These authors contributed equally to this work.}

\author[1]{\fnm{Guanzhong}~\sur{Wang}}
\equalcont{These authors contributed equally to this work.}

\author[1]{\fnm{Nick}~\sur{van~Loo}}
\equalcont{These authors contributed equally to this work.}

\author[1]{\fnm{Chun-Xiao}~\sur{Liu}}

\author[1]{\fnm{Grzegorz~P.}~\sur{Mazur}}

\author[1]{\fnm{Alberto}~\sur{Bordin}}

\author[1]{\fnm{Sebastiaan~L.~D.}~\sur{ten~Haaf}}

\author[1]{\fnm{Ji-Yin}~\sur{Wang}}

\author[1]{\fnm{David}~\sur{van~Driel}}

\author[1]{\fnm{Francesco}~\sur{Zatelli}}

\author[1]{\fnm{Xiang}~\sur{Li}}

\author[1]{\fnm{Filip~K.}~\sur{Malinowski}}

\author[2]{\fnm{Sasa}~\sur{Gazibegovic}}

\author[2]{\fnm{Ghada}~\sur{Badawy}}

\author[2]{\fnm{Erik~P.~A.~M.}~\sur{Bakkers}}

\author[1]{\fnm{Michael}~\sur{Wimmer}}

\author[1]{\fnm{Leo~P.}~\sur{Kouwenhoven}}

\affil[1]{\orgdiv{QuTech and Kavli Institute of NanoScience}, \orgname{Delft University of Technology}, \postcode{2600 GA} \orgaddress{\city{Delft}, \country{The Netherlands}}}

\affil[2]{\orgdiv{Department of Applied Physics}, \orgname{Eindhoven University of Technology}, \postcode{5600 MB} \orgaddress{\city{Eindhoven}, \country{The Netherlands}}}

\abstract{
Majorana bound states constitute one of the simplest examples of emergent non-Abelian excitations in condensed matter physics. 
A toy model proposed by Kitaev shows that such states can arise at the ends of a spinless $p$-wave superconducting chain~\cite{Kitaev.2001}. 
Practical proposals for its realization~\cite{Sau.2012, Leijnse.2012} require coupling  neighboring quantum dots in a chain via both electron tunneling and crossed Andreev reflection~\cite{Recher.2001}. 
While both processes have been observed in semiconducting nanowires and carbon nanotubes~\cite{hofstetter2009cooper,herrmann2010carbon,das2012high-efficiency,schindele2012near-unity}, crossed-Andreev interaction was neither easily tunable nor strong enough to induce coherent hybridization of dot states.
Here we demonstrate the simultaneous presence of all necessary ingredients for an artificial Kitaev chain: two spin-polarized quantum dots in an InSb nanowire strongly coupled by both elastic co-tunneling and crossed Andreev reflection. 
We fine-tune this system to a sweet spot where a pair of Poor Man's Majorana states is predicted to appear. At this sweet spot, the transport characteristics satisfy the theoretical predictions for such a system, including pairwise correlation, zero charge and stability against local perturbations.
While the simple system presented here can be scaled to simulate a full Kitaev chain with an emergent topological order, it can also be used imminently to explore relevant physics related to non-Abelian anyons. 
}

\maketitle

Engineering Majorana bound states in condensed matter systems is an intensively pursued goal, both for their exotic non-Abelian exchange statistics and for potential applications in building topologically protected qubits~\cite{Kitaev.2001, Nayak.2008, Kitaev.2003}. 
The most investigated experimental approach looks for Majorana states at the boundaries of topological superconducting materials, made of hybrid semiconducting-superconducting heterostructures~\cite{Mourik.2012,Deng2016,Fornieri2019,ren2019topological,Vaitieknas2020}.
However, the widely-relied-upon signature of Majorana states, zero-bias conductance peaks, is by itself unable to distinguish topological Majorana states from other trivial zero-energy states induced by disorder and smooth gate potentials~\cite{Kells2012,Prada2012,Pikulin2012,Liu2017andreev,Vuik.2019, Pan.2020.Physical}.
Both problems disrupting the formation or detection of a topological phase originate from a lack of control over the microscopic details of the electron potential landscape in these heterostructure devices.

In this work, we realize a minimal Kitaev chain~\cite{Kitaev.2001} using two quantum dots (QDs) coupled via a short superconducting-semiconducting hybrid~\cite{Sau.2012}. 
By controlling the electrostatic potential on each of these three elements, we overcome the challenge imposed by random disorder potentials. 
At a fine-tuned sweet spot where Majorana states are predicted to appear, we observe end-to-end correlated conductance that signals emergent Majorana properties such as zero charge and robustness against local perturbations.
We note that these Majorana states in a minimal Kitaev chain are \emph{not} topologically protected and have been dubbed ``Poor Man's Majorana" (PMM) states~\cite{Leijnse.2012}.

\begin{figure}[h]
\centering
\includegraphics[width=\textwidth]{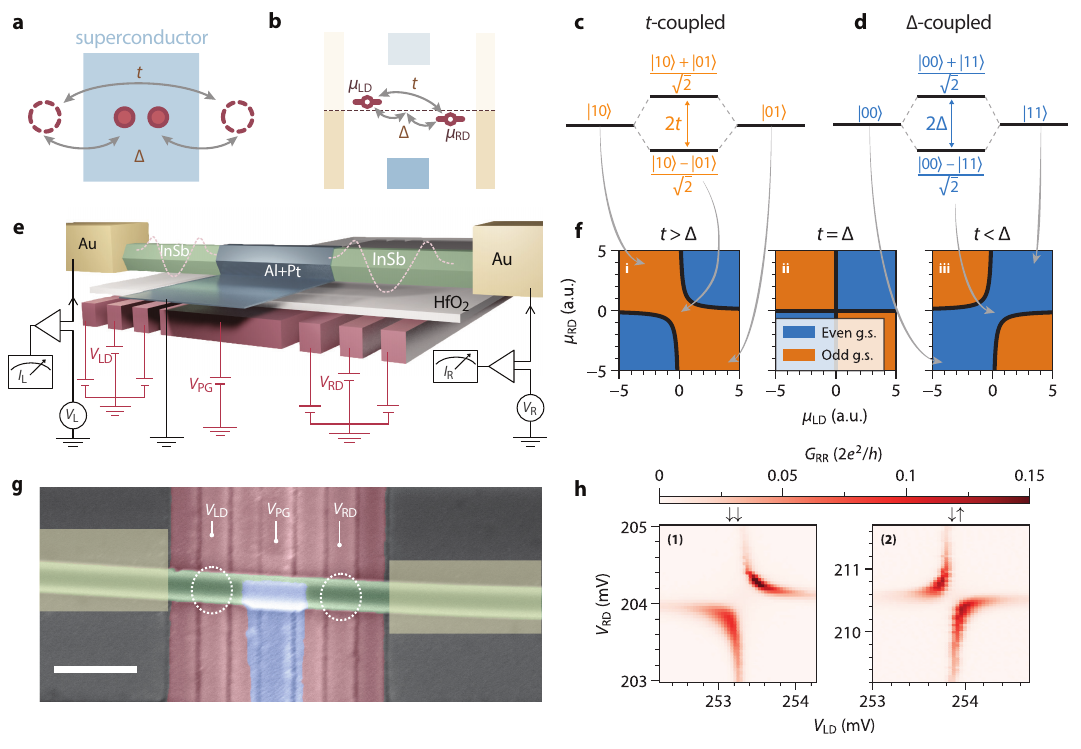}
\caption{\textbf{Coupling quantum dots through elastic co-tunneling (ECT) and crossed Andreev reflection (CAR).} \textbf{a.} Illustration of the basic ingredients of a Kitaev chain: two QDs simultaneously coupled via ECT with amplitude $t$ and via CAR with amplitude $\Delta$ through the superconductor in between. 
\textbf{b.} Energy diagram of a minimal Kitaev chain. Two QDs with gate-controlled chemical potentials are coupled via both ECT and CAR. The two ohmic leads enable transport measurements from both sides.
\textbf{c.} Energy diagram showing that coupling the $\ket{01}$ and $\ket{10}$ states via ECT leads to a bonding state $(\ket{10}-\ket{01})/\sqrt{2}$ and anti-bonding state $(\ket{10}+\ket{01})/\sqrt{2}$.
\textbf{d.} Same showing how CAR couples $\ket{00}$ and $\ket{11}$ to form the bonding state $(\ket{00}-\ket{11})/\sqrt{2}$ and anti-bonding state $(\ket{00}+\ket{11})/\sqrt{2}$ .
\textbf{e.} Illustration of the N-QD-S-QD-N device and the measurement circuit. Dashed potentials indicate QDs defined in the nanowire by finger gates.
\textbf{f.} Charge stability diagram of the coupled-QD system, in the cases of $t > \Delta $ (i), $t = \Delta $ (ii)  and $t < \Delta$ (iii). Blue marks regions in the ($\muL,\muR$) plane where the ground state is even and orange where the ground state is odd. 
\textbf{g.} False-colored scanning electron microscopy image of the device, prior to the fabrication of the normal leads. InSb nanowire is colored green. QDs are defined by bottom finger gates (in red) and their locations are circled. The gates controlling the two QD chemical potentials are labeled by their voltages, $\VLD$ and $\VRD$. The central thin Al/Pt film, in blue, is grounded. The proximitized nanowire underneath is gated by $\VPG$. Two Cr/Au contacts are marked by yellow boxes. 
The scale bar is \SI{300}{nm}.
\textbf{h.} Right-side zero-bias local conductance $\GRR$ in the $(\VLD,\VRD)$ plane when the system is tuned to $t > \Delta $ (1) and $t < \Delta$ (2).  The arrows mark the spin polarization of the QD levels. The DC bias voltages are kept at zero, $\VL=\VR=0$, and an AC excitation of \SI{6}{\micro V} RMS is applied on the right side.}
\label{fig:pmm-fig1}
\end{figure}

The elementary building block of the Kitaev chain is a pair of spinless electronic sites coupled simultaneously by two mechanisms: elastic co-tunneling (ECT) and crossed Andreev reflection (CAR). 
Both processes are depicted in \cref{fig:pmm-fig1}a. 
ECT involves a single electron hopping between two sites with an amplitude $t$. 
CAR refers to two electrons from both sites tunneling back and forth into a common superconductor with an amplitude $\Delta$ (not to be confused with the superconducting gap size), forming and splitting Cooper-pairs~\cite{Recher.2001}.
To create the two-site Kitaev chain, we utilize two spin-polarized QDs where only one orbital level in each dot is available for transport. 
In the absence of tunneling between the QDs, the system is characterized by a well-defined charge state on each QD: $\ket{\nL\ \nR}$, where $\nL,\nR \in \{0,1\}$ are occupations of the left and right QD levels. 
The charge on each QD depends only on its electrochemical potential $\muL$ or $\muR$, schematically shown in \cref{fig:pmm-fig1}b. 

In the presence of inter-dot coupling, the eigenstates of the combined system become superpositions of the charge states. 
ECT couples $\ket{10}$ and $\ket{01}$, resulting in two eigenstates of the form $ \alpha \ket{10} - \beta \ket{01}$ (\cref{fig:pmm-fig1}c), both with odd combined charge parity. 
These two bonding and anti-bonding states differ in energy by $2t$ when both QDs are at their charge degeneracy, i.e., $\muL=\muR=0$.
Analogously, CAR couples the two even states $\ket{00}$ and $\ket{11}$ to produce bonding and anti-bonding eigenstates of the form $u\ket{00} - v\ket{11}$, preserving the even parity of the original states. 
These states differ in energy by $2\Delta$ when $\muL=\muR=0$ (\cref{fig:pmm-fig1}d). 
If the amplitude of ECT is stronger than CAR ($t>\Delta$), the odd bonding state has lower energy than the even bonding state near the joint charge degeneracy $\muL=\muR=0$ (see Methods for details). 
The system thus features an odd ground state in a wider range of QD potentials, leading to a charge stability diagram shown in \cref{fig:pmm-fig1}f(i)~\cite{Wiel.2003erp}. 
The opposite case of CAR dominating over ECT, i.e., $t<\Delta$, leads to a charge stability diagram shown in \cref{fig:pmm-fig1}f(iii), where the even ground state is more prominent. 
Fine-tuning the system such that $t=\Delta$ equalizes the two avoided crossings, inducing an even-odd degenerate ground state at $\muL=\muR=0$ (\cref{fig:pmm-fig1}f(ii)).  
This degeneracy gives rise to two spatially separated PMMs, each localized at one QD~\cite{Leijnse.2012}.

\cref{fig:pmm-fig1}e illustrates our coupled QD system and the electronic measurement circuit. 
An InSb nanowire is contacted on two sides by two Cr/Au normal leads (N). 
A \SI{200}{nm}-wide superconducting lead (S) made of a thin Al/Pt film covering the nanowire is grounded and proximitizes the central semiconducting segment. 
The chemical potential of the proximitized semiconductor can be tuned by gate voltage $\VPG$. 
This hybrid segment shows a hard superconducting gap accompanied by discrete, gate-tunable Andreev bound states (\cref{fig:pmm-ED_dotchar}).
Two QDs are defined by finger gates underneath the nanowire.  
Their chemical potentials $\muL, \muR$ are linearly tuned by voltages on the corresponding gates $\VLD,\VRD$. 
Bias voltages on the two N leads, $\VL, \VR$, are applied independently and currents through them, $\IL, \IR$, are measured separately.
Transport characterization shows charging energies of \SI{1.8}{meV} on the left QD and \SI{2.3}{meV} on the right (\cref{fig:pmm-ED_dotchar}).  
Standard DC+AC lock-in technique allows measurement of the full conductance matrix:
\begin{equation}
G = \begin{pmatrix}
\GLL & \GLR \\
\GRL & \GRR
\end{pmatrix} = \begin{pmatrix}
\frac{\mathrm{d} \IL}{\mathrm{d} \VL} & \frac{\mathrm{d} \IL}{\mathrm{d} \VR} \\
\frac{\mathrm{d} \IR }{ \mathrm{d} \VL} & \frac{\mathrm{d} \IR}{\mathrm{d} \VR}
\end{pmatrix}.
\end{equation}
Measurements were conducted in a dilution refrigerator in the presence of a magnetic field $B = \SI{200}{mT}$ applied approximately along the nanowire axis. 
The combination of Zeeman splitting $E_\mathrm{Z}$ and orbital level spacing allows single-electron QD transitions to be spin-polarized. 
Two neighbouring Coulomb resonances correspond to opposite spin orientations, enabling the QD spins to be either parallel ($\uparrow \uparrow$ and $\downarrow\downarrow$) or anti-parallel ($\uparrow\downarrow$ and $\downarrow\uparrow$).
We report on two devices, A in the main text and B in Extended Data (\cref{fig:pmm-ED_device_2} and \cref{fig:pmm-ED_devB_biasGates}).
A scanning electron microscope image of Device~A is shown in \cref{fig:pmm-fig1}g.

Transport measurements are used to characterize the charge stability diagram of the system.
In \cref{fig:pmm-fig1}h(1), we show $\GRR$ as a function of QD voltages $\VLD, \VRD$ when both QDs are set to spin-down ($\downarrow \downarrow$).
The measured charge stability diagram shows avoided crossing which indicates the dominance of ECT.  
In \cref{fig:pmm-fig1}h(2), we change the spin configuration to $\downarrow \uparrow$.  
The charge stability diagram now develops the avoided crossing of the opposite orientation, indicating the dominance of CAR for QDs with anti-parallel spins.
This is, to our knowledge, the first verification of the prediction that spatially separated QDs can coherently hybridize via CAR coupling to a superconductor~\cite{choi2000spin}. 
Thus, we have introduced all the necessary ingredients for a two-site Kitaev chain. 

\section{Tuning the relative strength of CAR and ECT}

\begin{figure}[h]
\centering
\includegraphics[width=\textwidth]{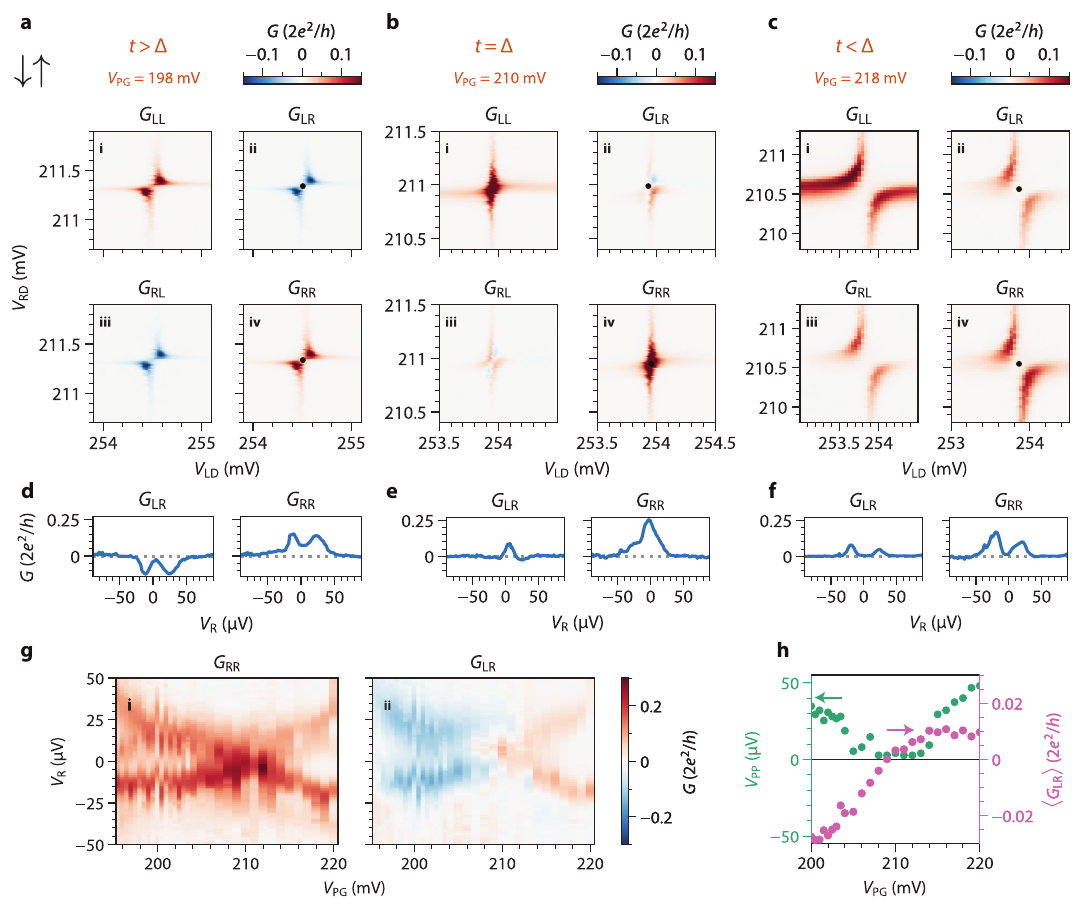}
\caption{\textbf{Tuning the relative strength of CAR and ECT for the $\downarrow\uparrow$ spin configuration.} \textbf{a-c.} Conductance matrices measured with  $\VPG = (198, 210, 218)~\si{mV}$, respectively.
\textbf{d-f.} $\GLR$ and $\GRR$ as functions of $\VR$ when $\VLD, \VRD$ are set to the center of each charge stability diagram in panels a to c, indicated by the black dots in the corresponding panels above them.
\textbf{g.} Local ($\GRR$) and nonlocal ($\GLR$) conductance as a function of $\VR$ and $\VPG$ while keeping $\muL \approx \muR \approx 0$, showing the continuous crossover from $t>\Delta$ to $t<\Delta$. 
\textbf{h.} Green dots: peak-to-peak distance ($V_\textrm{PP}$) between the positive- and negative-bias segments of $\GRR$, showing the closing and re-opening of QD avoided crossings.
Purple dots: average $\GLR$ ($\langle \GLR \rangle$) as a function of $\VPG$, showing a change in the sign of the nonlocal conductance.}
\label{fig:pmm-fig2}
\end{figure}

Majorana states in long Kitaev chains are present under a wide range of parameters due to topological protection~\cite{Kitaev.2001}. 
Strikingly, even a chain consisting of only two sites can host a pair of PMMs despite a lack of topological protection, if the fine-tuned sweet spot $t = \Delta$ and $\muL=\muR=0$ can be achieved~\cite{Leijnse.2012}. 
This, however, is made challenging by the above-mentioned requirement to have both QDs spin-polarized.
If spin is conserved, ECT can only take place between QDs with $\downarrow \downarrow$ or $\uparrow \uparrow$ spins, while CAR is only allowed for $\uparrow \downarrow$ and $\downarrow \uparrow$.
Rashba spin-orbit coupling in InSb nanowires solves this dilemma~\cite{Sau.2012,Liu.2022,wang2022Singlet}, allowing finite ECT even in anti-parallel spin configurations and CAR between QDs with equal spins.

A further challenge is to make the two coupling strengths equal for a given spin combination. 
Refs.~\cite{Liu.2022,wang2022Singlet,bordin2022controlled} show that both CAR and ECT in our device are virtual transitions through intermediate Andreev bound states residing in the short InSb segment underneath the superconducting film. 
Thus, varying $\VPG$ changes the energy and wavefunction of said Andreev bound states and thereby $t,\Delta$.
We search for the $\VPG$ range over which $\Delta$ changes differently than $t$ and look for a crossover in the type of charge stability diagrams.

\cref{fig:pmm-fig2}a-c shows the resulting charge stability diagrams for the $\downarrow\uparrow$ spin configuration at different values of $\VPG$. 
The conductance matrix $G(\VL = 0, \VR = 0)$ at $\VPG = \SI{198}{mV}$ is shown in \cref{fig:pmm-fig2}a. The local conductance on both sides, $\GLL$ and $\GRR$, exhibit level repulsion indicative of $t > \Delta$. 
We emphasize that ECT can become stronger than CAR even though the spins of the two QD transitions are anti-parallel due to the electric gating mentioned above. 
The dominance of ECT over CAR can also be seen in the negative sign of the nonlocal conductance, $\GLR$ and $\GRL$.
During ECT, an electron enters the system through one dot and exits through the other, resulting in negative nonlocal conductance.
CAR, in contrast, causes two electrons to enter or leave both dots simultaneously, producing positive nonlocal conductance~\cite{Beckmann2006}.
The residual finite conductance in the center of the charge stability diagram can be attributed to level broadening due to finite temperature and dot-lead coupling (see \cref{fig:pmm-ED_broadening_CSD}).
In \cref{fig:pmm-fig2}d, we show the conductance spectrum measured as a function of $\VR$, with $\VLD$ and $\VRD$ tuned to $\muL \approx \muR \approx 0$ (black dots in panels~c(ii, iv)). 
A pair of conductance peaks or dips is visible on either side of zero energy.

\cref{fig:pmm-fig2}c shows $G$ at $\VPG = \SI{218}{mV}$ (the $\GRR$ component is also used for \cref{fig:pmm-fig1}h(2)). 
Here, all the elements of $G$ exhibit CAR-type avoided crossings.
The spectrum shown in panel~f, obtained at the joint charge degeneracy point (black dots in panels~c(ii, iv)), similarly has two conductance peaks surrounding zero energy.
The measured nonlocal conductance is positive as predicted for CAR. 
The existence of both $t>\Delta$ and $t<\Delta$ regimes, together with continuous gate tunability, allows us to approach the $t \approx \Delta$ sweet spot. 
This is shown in panel~b, taken with $\VPG = \SI{210}{mV}$. 
Here, $\GRR$ and $\GLL$ exhibit no avoided crossing while $\GLR$ and $\GRL$ fluctuate around zero, confirming that CAR and ECT are in balance. 
Accordingly, the spectrum in panel~e confirms the even and odd ground states are degenerate and transport can occur at zero excitation energy via the appearance of a zero-bias conductance peak.
The crossover from the $t>\Delta$ regime to the $t<\Delta$ regime can be seen across multiple QD resonances (\cref{fig:pmm-ED_multi_res}).

To show that gate-tuning of the $t/\Delta$ ratio is indeed continuous, we repeat charge stability diagram measurements (\cref{fig:pmm-ED_VPG_DU}) and bias spectroscopy at more  $\VPG$ values.
As before, each bias sweep is conducted while keeping both QDs at charge degeneracy. 
\cref{fig:pmm-fig2}g shows the resulting composite plot of $\GRR$ (i) and $\GLR$ (ii) vs bias voltage and $\VPG$. 
The X-shaped conductance feature indicates a continuous evolution of the excitation energy, with a linear zero-energy crossing agreeing with predictions in Ref.~\cite{Leijnse.2012}. 
Following analysis described in Methods, we extract the peak spacing and average nonlocal conductance in \cref{fig:pmm-fig2}h in order to visualize the continuous crossover from $t>\Delta$ to $t<\Delta$.

\section{Poor Man's Majorana sweet spot}

\begin{figure}[h]
\centering
\includegraphics[width=\textwidth]{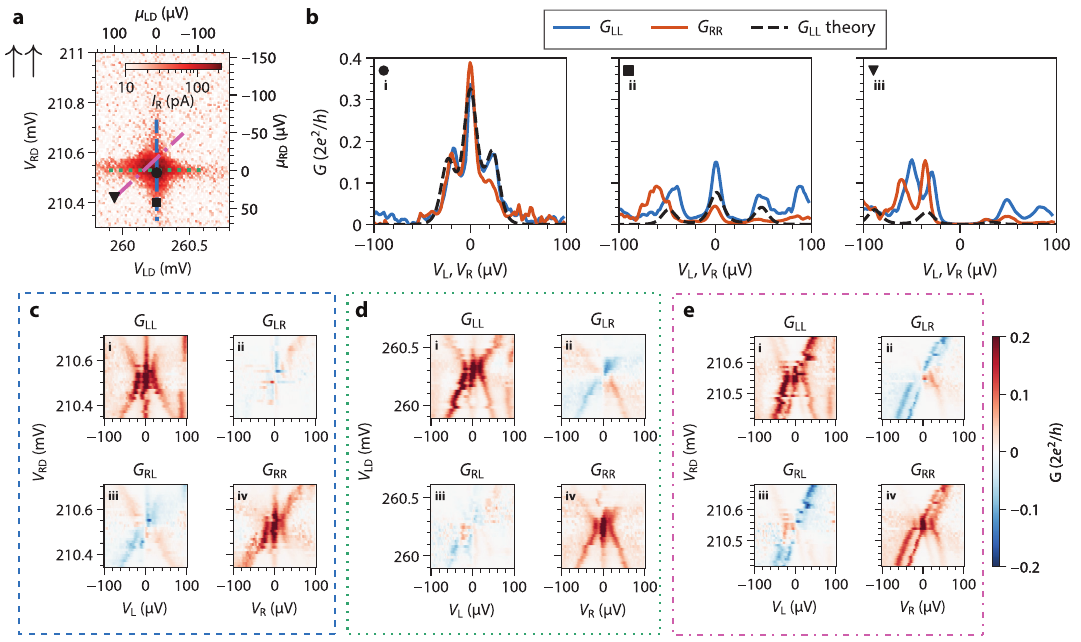}
\caption{\textbf{Conductance spectroscopy at the $t = \Delta$ sweet spot for the $\uparrow \uparrow$ spin configuration. } 
\textbf{a.} $\IR$ vs $\VLD,\VRD$ under $\VL= 0,\VR = \SI{10}{\micro V}$. The spectra in panel~b are taken at values of $\VLD,\VRD$ marked by corresponding symbols. The gate vs bias sweeps are taken along the dashed, dotted, dash-dot lines in panels~c,d,e respectively. Data are taken with fixed $\VPG = \SI{215.1}{mV}$.
\textbf{b.} Spectra taken under the values of $\VLD,\VRD$ marked in panel~a.
The dashed lines are theoretical curves calculated with $t=\Delta = \SI{12}{\micro eV}$, $\Gamma_\mathrm{L} = \Gamma_\mathrm{R} = \SI{4}{\micro eV}$,  $T = \SI{45}{mK}$ and at QD energies converted from $\VLD, \VRD$ using measured lever arms (see Methods for details).
\textbf{c, d.} $G$ as a function of the applied bias and $\VRD$ (c) or $\VLD$ (d), taken along the paths indicated by the dashed blue line and the dotted green line in panel~a, respectively. 
\textbf{e.} $G$ as a function of the applied bias and along the diagonal indicated by the dashed-dotted pink line in panel~a. This diagonal represents $\SI{500}{\micro V}$ of change in $\VLD$ and $\SI{250}{\micro V}$ of change in $\VRD$. }
\label{fig:pmm-fig3}
\end{figure}

Next, we study the excitation spectrum in the vicinity of the $t=\Delta$ sweet spot. 
The predicted zero-temperature experimental signature of the PMMs is a pair of quantized zero-bias conductance peak on both sides of the devices.
These zero-bias peaks are persistent even when one of the QD levels deviates from charge degeneracy~\cite{Leijnse.2012}.
We focus on the $\uparrow \uparrow$ spin configuration since it exhibits higher $t,\Delta$ values when they are equal (see \cref{fig:pmm-ED_VPG_UU}). 
\cref{fig:pmm-fig3}a shows the charge stability diagram measured via $\IR$ under fixed $\VL=0, \VR = \SI{10}{\micro V}$. 
No level repulsion is visible, indicating $t\approx \Delta$. 
Panel~b(i) shows the excitation spectrum when both dots are at charge degeneracy. 
The spectra on both sides show zero-bias peaks accompanied by two side peaks. 
The values of $t,\Delta$ can be read directly from the position of the side peaks, which correspond to the anti-bonding excited states at energy $2t = 2\Delta \approx \SI{25}{\micro eV}$. 
The height of the observed zero-bias peaks is 0.3 to $0.4\times2e^2/h$, likely owing to a combination of tunnel broadening and finite electron temperature (\cref{fig:pmm-ED_T_extra}). 
\cref{fig:pmm-fig3}b(ii) shows the spectrum when the right QD is moved away from charge degeneracy while $\muL$ is kept at 0. 
The zero-bias peaks persist on both sides of the device, as expected for a PMM state. 
In contrast, tuning both dots away from charge degeneracy, shown in \cref{fig:pmm-fig3}b(iii), splits the zero-bias peaks.

In \cref{fig:pmm-fig3}c,d, we show the evolution of the spectrum when varying $\VRD$ and $\VLD$, respectively.
The vertical feature appearing in both $\GLL$ and $\GRR$ shows correlated zero-bias peaks in both QDs, which persist when one QD potential departs from zero. 
This crucial observation demonstrates the robustness of PMMs against local perturbations. 
The excited states disperse in agreement with the theoretical predictions~\cite{Leijnse.2012}. 
Nonlocal conductance, on the other hand, reflects the local charge character of a bound state on the side where current is measured~\cite{gramich2017andreev,Danon.2020,Menard.2020}.
Near-zero values of $\GLR$ in panel~c and $\GRL$ in panel~d are consistent with the prediction that the PMM mode on the unperturbed side remains an equal superposition of an electron and a hole and therefore chargeless. 

Finally, when varying the chemical potential of both dots simultaneously (panel~e), we see that the zero-bias peaks split away from zero energy. 
This splitting is not linear, in contrast to the case when $\Delta \ne t$ (see \cref{fig:pmm-ED_NL_ABS}).
The profile of the peak splitting is consistent with the predicted quadratic protection of PMMs against chemical potential fluctuations~\cite{Leijnse.2012}.
This quadratic protection is expected to develop into topological protection in a long-enough Kitaev chain~\cite{Sau.2012}.

\section{Discussion}

\begin{figure}[h]
\centering
\includegraphics[width=\textwidth]{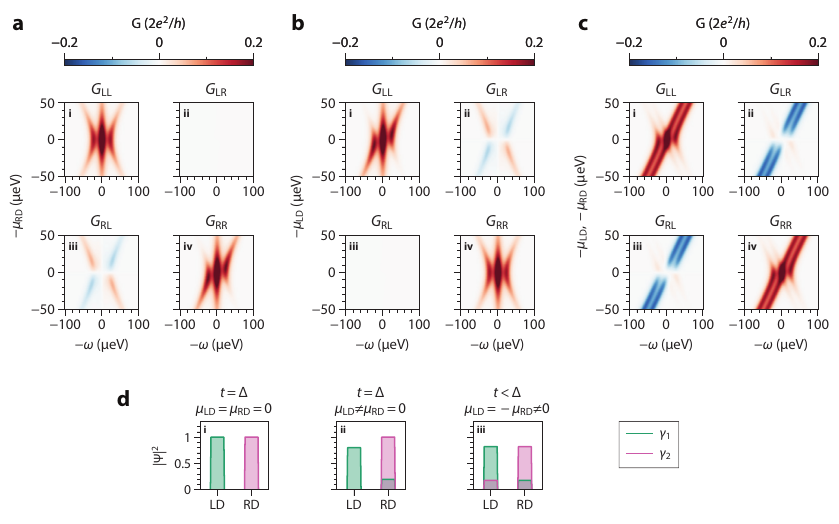}
\caption{\textbf{Calculated conductance and Majorana localization.} 
\textbf{a.} Numerically calculated $G$ as a function of energy $\omega$ and $\muR$ at the $t = \Delta$ sweet spot.  
\textbf{b.} Numerically calculated $G$ as a function of $\omega$ and $\muL$ at the $t = \Delta$ sweet spot. 
\textbf{c.} Numerically calculated $G$ as a function of $\omega$ and $\muR,\muL$ along the diagonal corresponding to $\muL = \muR$ at $t = \Delta$. All of the numerical curves use the same value of $t,\Delta,\Gamma_\mathrm{L},\Gamma_\mathrm{R}$ as those in Fig.3.
\textbf{d.} Illutrations of the localization of two zero-energy solutions for the following set of parameters: $t=\Delta$, $\muL=\muR=0$ (sub-panel~i), $t=\Delta$, $\muR=0$, $\muL>0$ (sub-panel~ii), $t<\Delta$, $\muL=-\muR=\sqrt{\Delta^2-t^2}$ (sub-panel~iii).}
\label{fig:pmm-fig4}
\end{figure}

To facilitate comparison with data, we develop a transport model (see Methods) and plot in \cref{fig:pmm-fig4}a-c the calculated conductance matrices as functions of excitation energy, $\omega$, vs $\muR$ (panel~a), $\muL$ (panel~b), and $\mu \equiv \muL = \muR$ (panel~c). 
These conditions are an idealization of those in \cref{fig:pmm-fig3} (a more realistic simulation of the experimental conditions is presented in \cref{fig:pmm-ED_alt_fig4}). 
The numerical simulations capture the main features appearing in the experiments discussed above.  

Particle-hole symmetry ensures that zero-energy excitations in this system always come in pairs. 
These excitations can extend over both QDs or be confined to one of them. 
In \cref{fig:pmm-fig4}d we show the calculated spatial extent of the zero-energy excitations for three scenarios. 
The first, in \cref{fig:pmm-fig4}d(i), illustrates \cref{fig:pmm-fig3}b(i) and shows that the sweet-spot zero-energy solutions are two PMMs, each localized on a different QD. 
The second scenario in \cref{fig:pmm-fig4}d(ii), illustrating \cref{fig:pmm-fig3}b(ii), is varying $\muL$ while keeping $\muR=0$.
This causes some of the wavefunction localized on the perturbed left side, $\gamma_1$, to leak into the right QD. 
Since the right-side $\gamma_2$ excitation has no weight on the left, it does not respond to this perturbation and remains fully localized on the right QD.
As the theory confirms~\cite{Leijnse.2012}, it stays a zero-energy PMM state. 
Since Majorana excitations always come in pairs, the excitation on the left QD must also remain at zero energy. 
This provides an intuitive understanding of the remarkable stability of the zero-energy modes at the sweet spot in \cref{fig:pmm-fig3}c,d when moving one of the QDs' chemical potentials away from zero.
Finally, zero-energy solutions can be found away from the sweet spot, $t \ne \Delta$, as illustrated in \cref{fig:pmm-fig4}d(iii). 
These zero-energy states are only found when both QDs are off-resonance and none of them are localized Majorana states, extending over both QDs and exhibiting no gate stability. 
Measurements under these conditions are shown in \cref{fig:pmm-ED_NL_ABS}, where zero-energy states can be found in a variety of gate settings (panels~a,~c therein).

\section{Conclusion}

In summary, we realize a minimal Kitaev chain where two QDs in an InSb nanowire are separated by a hybrid semiconducting-superconducting segment. 
Compared to past works, our approach solves three challenges: strong hybridization of QDs via CAR, simultaneous coupling of two single spins via both ECT and CAR, and continuous tuning of the coupling amplitudes. 
This is made possible by the two QDs as well as the middle Andreev bound state mediating their couplings all being discrete, gate-tunable quantum states.
The result is the creation of a new type of nonlocal states that host Majorana-type excitations at a fine-tuned sweet spot. 
The zero-bias peaks at this spot are robust against variations of the chemical potential of one QD and quadratically protected against simultaneous perturbations of both.
This discrete and tunable way of assembling Kitaev chains shows good agreement between theory and experiment by avoiding the most concerning problems affecting the continuous nanowire experiments: disorder, smooth gate potentials and multi-subband occupation~\cite{Pan.2021.KitaevVsNanowire}.  
The QD-S-QD platform discussed here opens up a new frontier to the study of Majorana physics. In the long term, this approach can generate topologically protected Majorana states in longer chains~\cite{Sau.2012}.
A shorter term approach is to use PMMs as an immediate playground to study fundamental non-Abelian statistics, e.g., by fusing neighboring PMMs in a device with two such copies.

\section{Methods}

\subsection{Device fabrication}

The nanowire hybrid devices presented in this work were fabricated on pre-patterned substrates, using the shadow-wall lithography technique described in Refs.~\cite{Heedt:2021_NC,Borsoi:2021_AFM}. 
Nanowires were deposited onto the substrates using an optical micro-manipulator setup. 
\SI{8}{nm} of Al was grown at a mix of 15$^\circ$ and 45$^\circ$ angles with respect to the substrate. 
Subsequently, Device~A was coated with \SI{2}{\angstrom} of Pt grown at 30$^\circ$. No Pt was deposited for Device~B. 
Finally, all devices were capped with \SI{20}{nm} of evaporated AlO$_x$. 
Details of the substrate fabrication, the surface treatment of the nanowires, the growth conditions of the superconductor, the thickness calibration of the Pt coating and the ex-situ fabrication of the ohmic contacts can be found in Ref.~\cite{Mazur.2022}.
Devices A and B also slightly differ in the length of the hybrid segment: \SI{180}{nm} for A and \SI{150}{nm} for B. 

\subsection{Transport measurement and data processing}

We have fabricated and measured six devices with similar geometry. 
Two of them showed strong hybridization of the QD states by means of CAR and ECT. 
We report on the detailed measurements of Device~A in the main text and show qualitatively similar measurements from Device~B in \cref{fig:pmm-ED_device_2} and \cref{fig:pmm-ED_devB_biasGates}.
All measurements on Device~A were done in a dilution refrigerator with base temperature \SI{7}{mK} at the cold plate and electron temperature of 40$\sim$\SI{50}{mK} at the sample, measured in a similiar setup using an NIS metallic tunnel junction. 
Unless otherwise mentioned, the measurements on Device A were conducted in the presence of a magnetic field of \SI{200}{mT} approximately oriented along the nanowire axis with a 3$^\circ$ offset. 
Device~B was measured similarly in another dilution refrigerator under $B=\SI{100}{mT}$ along the nanowire with 4$^\circ$ offset.

\cref{fig:pmm-fig1}e shows a schematic depiction of the electrical setup used to measure the devices. 
The middle segment of the InSb nanowire is covered by a thin Al shell, kept grounded throughout the experiment.  
On each side of the hybrid segment, we connect the normal leads to a current-to-voltage converter. 
The amplifiers on the left and right sides of the device are each biased through a digital-to-analog converter that applies DC and AC biases. 
The total series resistance of the voltage source and the current meter is less than \SI{100}{\ohm} for Device~A and \SI{1.11}{k\ohm} for Device~B. 
Voltage outputs of the current meter are read by digital multimeters and lock-in amplifiers. 
When DC voltage $\VL$ is applied, $\VR$ is kept grounded and vice versa. 
AC excitations are applied on each side of the device with different frequencies (\SI{17}{Hz} on the left and \SI{29}{Hz} on the right for Device~A, \SI{19}{Hz} on the left and \SI{29}{Hz} on the right for Device~B) and with amplitudes between 2 and \SI{6}{\micro V} RMS. 
In this manner, we measure the DC currents $\IL,\IR$ and the conductance matrix $G$ in response to applied voltages $\VL,\VR$ on the left and right N leads, respectively. 
The conductance matrix is corrected for voltage divider effects (see Ref.~\cite{Martinez2021} for details) taking into account the series resistance of sources and meters and in each fridge line (\SI{1.85}{k\ohm} for Device~A and \SI{2.5}{k\ohm} for Device~B), except for the right panel of \cref{fig:pmm-fig1}h and \cref{fig:pmm-fig2}d. 
There, the left half of the conductance matrix was not measured and correction is not possible. 
We verify that the series resistance is much smaller than device resistance and the voltage divider effect is never more than $\sim 10\%$ of the signal.

\subsection{Characterization of QDs and the hybrid segment}

To form the QDs described in the main text, we pinch off the finger gates next to the three ohmic leads, forming two tunnel barriers in each N-S junction. 
$\VLD$ and $\VRD$ applied on the middle finger gates on each side accumulate electrons in the QDs. 
We refer to the associated data repository for the raw gate voltage values used in each measurement.
See \cref{fig:pmm-ED_dotchar}a-f for results of the dot characterizations. 

Characterization of the spectrum in the hybrid segment is done using conventional tunnel spectroscopy. 
In each uncovered InSb segment, we open up the two finger gates next to the N lead and only lower the gate next to the hybrid to define a tunnel barrier. 
The results of the tunnel spectroscopy are shown in \cref{fig:pmm-ED_dotchar}g,h and the raw gate voltages are available in the data repository.

\subsection{Determination of QD spin polarization}

Control of the spin orientation of QD levels is done via selecting from the even vs odd charge degeneracy points following the method detailed in Ref.~\cite{Hanson.2007}.
At the charge transition between occupancy $2n$ and $2n+1$ ($n$ being an integer), the electron added to or removed from the QD is polarized to spin-down ($\downarrow$, lower in energy). 
The next level available for occupation, at the transition between $2n+1$ and $2n+2$ electrons, has the opposite polarization of spin-up ($\uparrow$, higher in energy). 
To ensure the spin polarization is complete, the experiment was conducted with $E_\mathrm{Z} \approx \SI{400}{\micro eV} > \lvert e\VL\rvert, \lvert e\VR \rvert$ (see \cref{fig:pmm-ED_dotchar} for determination of the spin configuration). 
In the experiment data, a change in the QD spin orientation is visible as a change in the range of $\VLD$ or $\VRD$.

\subsection{Controlling ECT and CAR via electric gating}

Ref.~\cite{Liu.2022} describes a theory of mediating CAR and ECT transitions between QDs via virtual hopping through an intermediate Andreev bound state.
Ref.~\cite{bordin2022controlled} experimentally verifies the applicability of this theory to our device.
To summarize the findings here, we consider two QDs both tunnel-coupled to a central Andreev bound state in the hybrid segment of the device.
The QDs have excitation energies lower than that of the Andreev bound state and thus transition between them is second-order.
The wavefunction of an Andreev bound state consists of a superposition of an electron part, $u$, and a hole part, $v$.
Both theory and experiment conclude that the values of $t$ and $\Delta$ depend strongly and differently on $u, v$. 
Specifically, CAR involves converting an incoming electron to an outgoing hole and thus depends on the values of $u$ and $v$ jointly as $\lvert uv \rvert^2$. 
ECT, however, occurs over two parallel channels (electron-to-electron and hole-to-hole) and its coupling strength depends on $u,v$ independently as $\lvert u^2-v^2 \rvert^2$. 
As the composition of $u,v$ is a function of the chemical potential of the middle Andreev bound state, the CAR to ECT ratio is strongly tunable by $\VPG$. 
We thus look for a range of $\VPG$ where Andreev bound states reside in the hybrid segment, making sure that the energies of these states are high enough so as not to hybridize with the QDs directly (\cref{fig:pmm-ED_dotchar}). 
Next, we sweep $\VPG$ to find the crossover point between $t$ and $\Delta$ as described in the main text.

\subsection{Additional details on the measurement of the coupled QD spectrum}

The measurement of the local and nonlocal conductance shown in \cref{fig:pmm-fig2}g was conducted in a series of steps. 
First, the value of $\VPG$ was set, and a charge stability diagram was measured as a function of $\VLD$ and $\VRD$. 
Representative examples of such diagrams are shown in \cref{fig:pmm-ED_VPG_DU}. 
Second, each charge stability diagram was inspected and the joint charge degeneracy point ($\muL=\muR=0$) was selected manually ($\VLD^0,\VRD^0$). 
Lastly, the values of $\VLD$ and $\VRD$ were set to those of the joint degeneracy point and the local and nonlocal conductance were measured as a function of $\VR$.

The continuous transition from $t>\Delta$ to $t<\Delta$ is visible in \cref{fig:pmm-fig2}g via both local and nonlocal conductance.
$\GRR$ shows that level repulsion splits the zero-energy resonance peaks both when $t > \Delta$ (lower values of $\VPG$) and when $t < \Delta $ (higher values of $\VPG$). 
The zero-bias peak is restored in the vicinity of $t = \Delta$, in agreement with theoretical predictions~\cite{Leijnse.2012}. 
The crossover is also apparent in the sign of $\GLR$, which changes from negative ($t > \Delta$) to positive ($t < \Delta $).

To better visualize the transition between the ECT- and CAR-dominated regimes, we extract $V_\textrm{PP}$, the separation between the conductance peaks under positive and negative bias voltages, and plot them as a function of $\VPG$ in \cref{fig:pmm-fig2}h. 
When tuning $\VPG$, the peak spacing decreases until the two peaks merge at $\VPG \approx \SI{210}{mV}$.  
Further increase of $\VPG$ leads to increasing $V_\textrm{PP}$. 
In addition, to observe the change in sign of the nonlocal conductance, we follow $\langle \GLR \rangle$, the value of $\GLR$ averaged over the bias voltage $\VR$ between $-100$ and \SI{100}{\micro V} at a given $\VPG$. 
We see that $\langle \GLR \rangle$ turns from negative to positive at $\VPG \approx \SI{210}{mV}$, in correspondence to a change in the dominant coupling mechanism.

\cref{fig:pmm-fig3}c-e presents measurements where the conductance was measured against applied biases along some paths within the charge stability diagram (panel~a). 
Prior to each of these measurements, a charge stability diagram was measured and inspected, based on which the relevant path in the $(\VLD,\VRD)$ plane was chosen. 
Following each bias spectroscopy measurement, another charge stability diagram was measured and compared to the one taken before to check for potential gate instability. 
In case of noticeable gate drifts between the two, the measurement was discarded and the process was repeated. 
The values of $\muL$ and $\muR$ required for theoretical curves appearing in panel~b were calculated by $\mu_i = \alpha_i (V_{i} - V_{i}^0)$ where $i=\mathrm{LD,RD}$ and $\alpha_i$ is the lever arm of the corresponding QD.
The discrepancy between the spectra measured with $\GLL$ and $\GRR$ likely results from gate instability, since they were not measured simultaneously. 
Finite remaining $\GLR$ in panel~c and $\GRL$ in panel~d most likely result from small deviations of $\muL,\muR$ from zero during these measurements.

\subsection{Model of the phase diagrams in \cref{fig:pmm-fig1}f}

To calculate the ground state phase diagram in \cref{fig:pmm-fig1}f, we write the Hamiltonian in the many-body picture, with the four basis states being $\ket{00},\ket{11}, \ket{10}, \ket{01}$:
\begin{equation}
H_\mathrm{mb}=
\begin{pmatrix}
0 & \Delta & 0 & 0 \\
\Delta & \varepsilon_L+\varepsilon_R & 0 & 0 \\
0 & 0 & \varepsilon_L & t \\
0 & 0 & t & \varepsilon_R
\end{pmatrix}
\end{equation}
in block-diagonalized form.
The two $2\times 2$ matrices yield the energy eigenvalues separately for the even and odd subspaces:
\begin{align}
E_{o,\pm} &=\frac{\varepsilon_L+\varepsilon_R}{2} \pm \sqrt{\left(\frac{\varepsilon_L-\varepsilon_R}{2} \right)^2 + t^2} \\
E_{e,\pm} &=\frac{\varepsilon_L+\varepsilon_R}{2} \pm \sqrt{\left(\frac{\varepsilon_L+\varepsilon_R}{2} \right)^2 + \Delta^2} 
\end{align}
The ground state phase transition occurs at the boundary $E_{o,-}=E_{e,-}$.
This is equivalent to 
\begin{equation}
    \varepsilon_L \varepsilon_R = t^2-\Delta^2
\end{equation}

\subsection{Transport model in \cref{fig:pmm-fig3} and \cref{fig:pmm-fig4}}

We describe in this section the model Hamiltonian of the minimal Kitaev chain and the method we use for calculating the differential conductance matrices when the Kitaev chain is tunnel-coupled to two external N leads.

The effective Bogoliubov-de-Gennes Hamiltonian of the double-QD system is
\begin{align}
H &= \varepsilon_L c\dg_L c_L + \varepsilon_R c\dg_R c_R 
+t c\dg_L c_R +t  c\dg_R c_L + \Delta c_L c_R + \Delta c\dg_R c\dg_L \nn
&=
\half 
\Psi\dg
\begin{pmatrix}
\varepsilon_L & t & 0 &-\Delta \\
t & \varepsilon_R & \Delta & 0 \\
0 & \Delta & -\varepsilon_L & -t \\
-\Delta & 0 & -t & -\varepsilon_R
\end{pmatrix}
\Psi,
\end{align}
where $\Psi = (c_L, c_R, c\dg_L, c\dg_R)\tp$ is the Nambu spinor, $\varepsilon_{L/R}$ is the level energy in dot-$L/R$ relative to the superconducting Fermi surface, $t$ and $\Delta$ are the ECT and CAR amplitudes.
Here we assume $t$ and $\Delta$ to be real without loss of generality \cite{Leijnse.2012}.
The presence of both $t$ and $\Delta$ in this Hamiltonian implies breaking spin conservation during QD-QD tunneling via either spin-orbit coupling (as done in the present experiment) or non-collinear magnetization between the two QDs (as proposed in \cite{Leijnse.2012}). Without one of them, equal-spin QDs cannot recombine into a Cooper pair, leading to vanishing $\Delta$, while opposite-spin QDs cannot support finite $t$. The exact values of $t$ and $\Delta$ depend on the spin-orbit coupling strength and we refer to Ref.~\cite{Liu.2022} for a detailed discussion.

To calculate the differential conductance for the double-QD system, we use the $S$-matrix method~\cite{datta2005quantum}.
In the wide-band limit, the $S$ matrix is  
\begin{align}
S(\omega) =
\begin{pmatrix}
s_{ee} & s_{eh} \\
s_{he} & s_{hh} 
\end{pmatrix}
=1 -  i W\dg \left( \omega - H + \half i W W\dg \right)^{-1} W,
\end{align}
where $ W =  \mathrm{diag} \{ \sqrt{\Gamma_L}, \sqrt{\Gamma_R}, -\sqrt{\Gamma_L}, -\sqrt{\Gamma_R} \}$ is the tunnel matrix, with $\Gamma_{\alpha} $ being the tunnel coupling strength between dot-$\alpha$ and lead-$\alpha$.
The zero-temperature differential conductance is given by
\begin{align}
G^{0}_{\alpha \beta}(\omega) = \mathrm{d} I _{\alpha} / \mathrm{d} V_{\beta} = \frac{e^2}{h} \left( \delta_{\alpha \beta}- \vert s^{\alpha \beta}_{ee}(\omega) \vert^2 + \vert s^{\alpha \beta}_{he}(\omega)\vert^2 \right),
\end{align}
where $\alpha,\beta=L/R$. 
Finite-temperature effect is included by a convolution between the zero-temperature conductance and the derivative of Fermi-Dirac distribution, i.e.,
\begin{align}
G^{T}(\omega) = \int dE  \frac{G^{0}(E)}{4k_BT \cosh^2[(E-\omega)/2k_BT]}.
\end{align}

The theoretical model presented above uses five input parameters to calculate the conductance matrix under given $\muL,\muR,\VL,\VR$. 
The input parameters are: $t,\Delta,\Gamma_L,\Gamma_R,T$. 
To choose the parameters in \cref{fig:pmm-fig3}b(i), we fix the temperature to the measured value $T = \SI{45}{mK}$ and make the simplification $t=\Delta$, $\Gamma\equiv\Gamma_L=\Gamma_R$. 
This results in only two free parameters $t,\Gamma$, which we manually choose and compare with data. 
While oversimplified, this approach allows us to obtain a reasonable match between theory and data taken at $\muL=\muR=0$ without the risk of overfitting.  
To obtain the other numerical curves shown in \cref{fig:pmm-fig3}, we keep the same choice of $t,\Gamma$ and vary $\muL,\muR,\VL,\VR$ along various paths in the parameter space. 
Similarly, to model the data shown in \cref{fig:pmm-ED_NL_ABS}, we keep $T = \SI{45}{mK}$ and $\Gamma$ the same as in \cref{fig:pmm-fig3}. 
The free parameters to be chosen are thus $t$ and $\Delta$.
The theory panels are obtained with the same $t,\Delta$, and only $\muL,\muR,\VL,\VR$ are varied in accordance with the experimental conditions.

Finally, we comment on the physical meaning of the theory predictions in \cref{fig:pmm-fig4}a-c.
Tuning $\muR$ leads to symmetric $\GLL$ and asymmetric $\GRR$, as well as zero $\GLR$ and finite $\GRL$ with an alternating pattern of positive and negative values. 
As discussed in the main text, these features, also seen in the measurements, stem from the local charge of the system: keeping $\muL=0$ maintains zero local charge on the left dot, while varying $\muR$ creates finite local charge on the right dot. 
The complementary picture appears when varying $\muL$ in panel~b. 
The asymmetry in both $\GLL$ and $\GRR$ and the negative nonlocal conductance when tuning simultaneously $\muL=\muR$ are also captured in the numerical simulation in panel~c.
We note that while there is a qualitative agreement between the features in \cref{fig:pmm-fig4}c and \cref{fig:pmm-fig3}e, they were obtained under nominally different conditions. 
As mentioned, the theoretical curve follows $\muL = \muR$, while the experimental curve was taken through a path along which $\VLD$ changed twice as much as $\VRD$, although the lever arms of both QDs are similar. 
In \cref{fig:pmm-fig4}c, we calculate the conductance along a path reproducing the experimental conditions. 
We speculate that the discrepancy between \cref{fig:pmm-fig3}e and \cref{fig:pmm-fig4}c could arise from some hybridization between the left QD and the superconducting segment as seen in \cref{fig:pmm-ED_dotchar}.

\subsection{Data Availability and Code Availability}
Raw data presented in this work, the data processing/plotting code and code used for the theory calculations are available at \url{https://doi.org/10.5281/zenodo.6594169}.

\clearpage

\renewcommand\thefigure{ED\arabic{figure}}
\setcounter{figure}{0}

\section{Extended data}


\begin{figure}[htbp]
\centering
\includegraphics[width=0.8\textwidth]{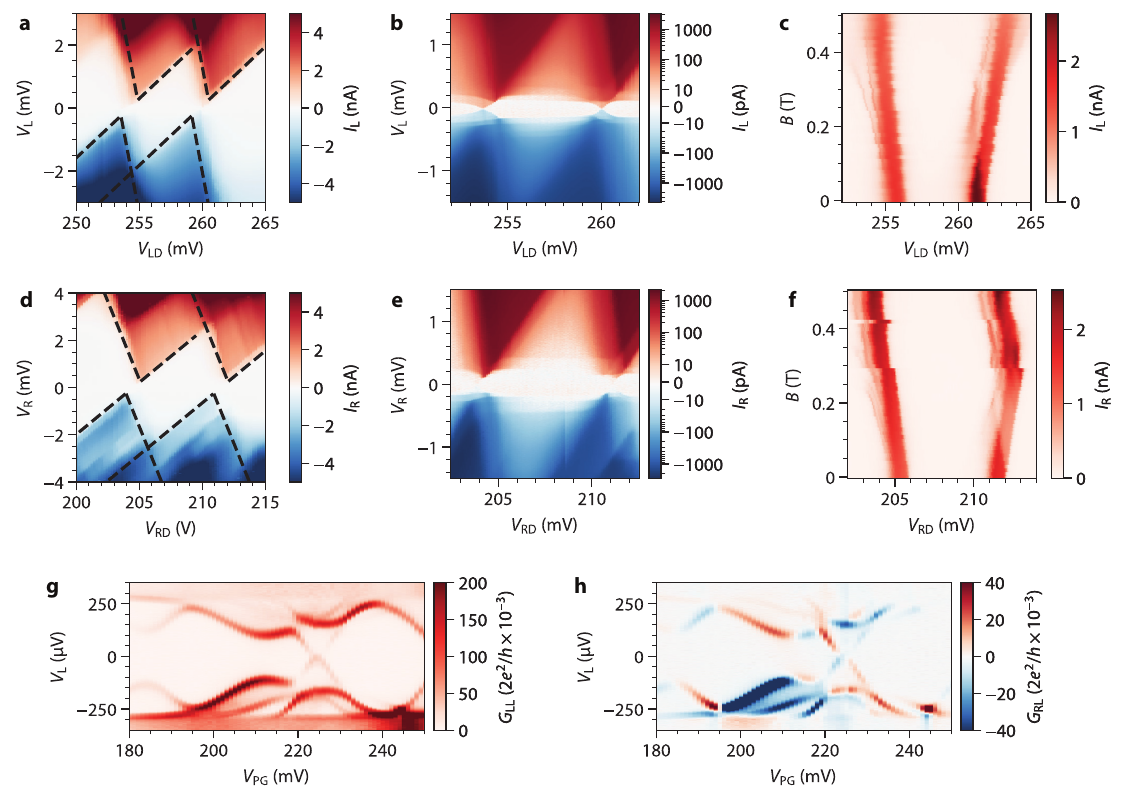}
\caption{\textbf{Characterization of the QDs.}
\textbf{a.} Coulomb blockade diamonds of the left QD when right QD is off resonance. $\IL$ is measured as a function of $\VL,\VLD$. The data is overlaid with a constant interaction model~\cite{Kouwenhoven2001} with \SI{1.8}{meV} charging energy and gate lever arm of 0.32.
\textbf{b.} A high-resolution scan of \textbf{a} with a symmetric-logarithmic color scale to show the presence of a small amount of Andreev current at sub-gap energies.
This is due to the left QD being weakly proximitized by local Andreev coupling to Al.
\textbf{c.} Field dependence of the Coulomb resonances. $\IL$ is measured as a function of $\VLD$ and $B$ with a constant $\VL = \SI{600}{\micro V}$. The resonances of opposite spin polarization evolve in opposite directions with a $g$-factor of $\sim 35$, translating to Zeeman energy of \SI{400}{\micro eV} at $B=\SI{200}{mT}$.
\textbf{d-f.} Characterization of the right QD, as described in the captions of panels~a-c.
Overlaid model in \textbf{d} has charging energy \SI{2.3}{meV} and gate lever arm of 0.33.
No sub-gap transport is detectable in \textbf{e}.
$B$ dispersion in \textbf{f} corresponds to $g=40$.
\textbf{g, h.} Bias spectroscopy results of the proximitized InSb segment under the thin Al/Pt film. 
$\IL,\IR$ are measured as a function of $\VL,\VPG$. 
$\GLL, \GRL$ are obtained by taking the numerical derivative of $\IL,\IR$ along the bias direction after applying a Savitzky-Golay filter of window length 15 and order 1.
The sub-gap spectrum reveals discrete, gate-dispersing Andreev bound states. The presence of nonlocal conductance correlated with the sub-gap states shows that these Andreev bound states extend throughout the entire hybrid segment, coupling to both left and right N leads~\cite{Menard.2020}.
Parts of this dataset are also presented in Ref.~\cite{Mazur.2022} (Reproduced under the terms of the CC-BY Creative Commons Attribution 4.0 International license (https://creativecommons.org/licenses/by/4.0). Copyright 2022, The Authors, published by Wiley-VCH).
}
\label{fig:pmm-ED_dotchar}
\end{figure}


\begin{figure}[htbp]
\centering
\includegraphics[width=0.5\textwidth]{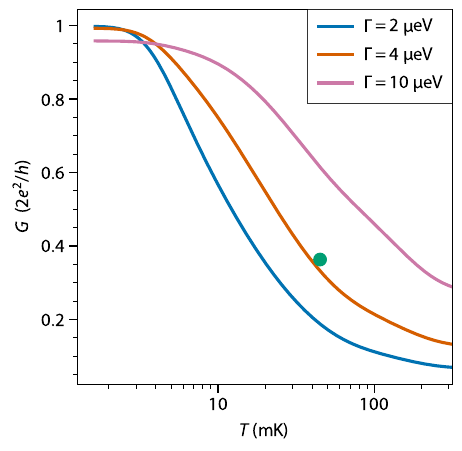}
\caption{
\textbf{Theoretical temperature dependence of the height of Majorana zero-bias conductance peaks.}
The height of the Majorana zero-bias peaks is only quantized to $2e^2/h$ at zero temperature.
At finite electron temperature $T$, the peak height is generally lower, with the exact value depending on $T$ and tunnel broadening $\Gamma_\mathrm{L}, \Gamma_\mathrm{R}$ due to coupling between QDs and N leads.
The local zero-bias conductance $\GLL$ at the sweet spot ($t=\Delta, \muL=\muR=0$) is calculated and shown in this plot as a function of $T$, using the parameters presented in \cref{fig:pmm-fig3}: $t = \Delta = \SI{12}{\micro eV}$.
Three curves are calculated assuming three different values of tunnel coupling $\Gamma = \Gamma_\mathrm{L} = \Gamma_\mathrm{R}$. 
The orange curve assumes a $\Gamma$ value that matches the experimentally observed peak width (both of the zero-bias peaks and of generic QD resonant peaks at other conductance features), showing that conductance approaching quantization would only be realized at electron temperatures  $<\SI{5}{mK}$, unattainable in our dilution refrigerator. The blue curve, calculated with lower $\Gamma = \SI{2}{\micro eV}$, shows even lower conductance. Increasing $\Gamma$ would not lead to conductance quantization either, since the zero-bias peaks would merge with the conductance peaks arising from the excited states (pink curve). The green dot marks the experimentally measured electron temperature and peak height (averaged between the values obtained on the left and right leads).}
\label{fig:pmm-ED_T_extra}
\end{figure}


\begin{figure}[htbp]
\centering
\includegraphics[width=0.9\textwidth]{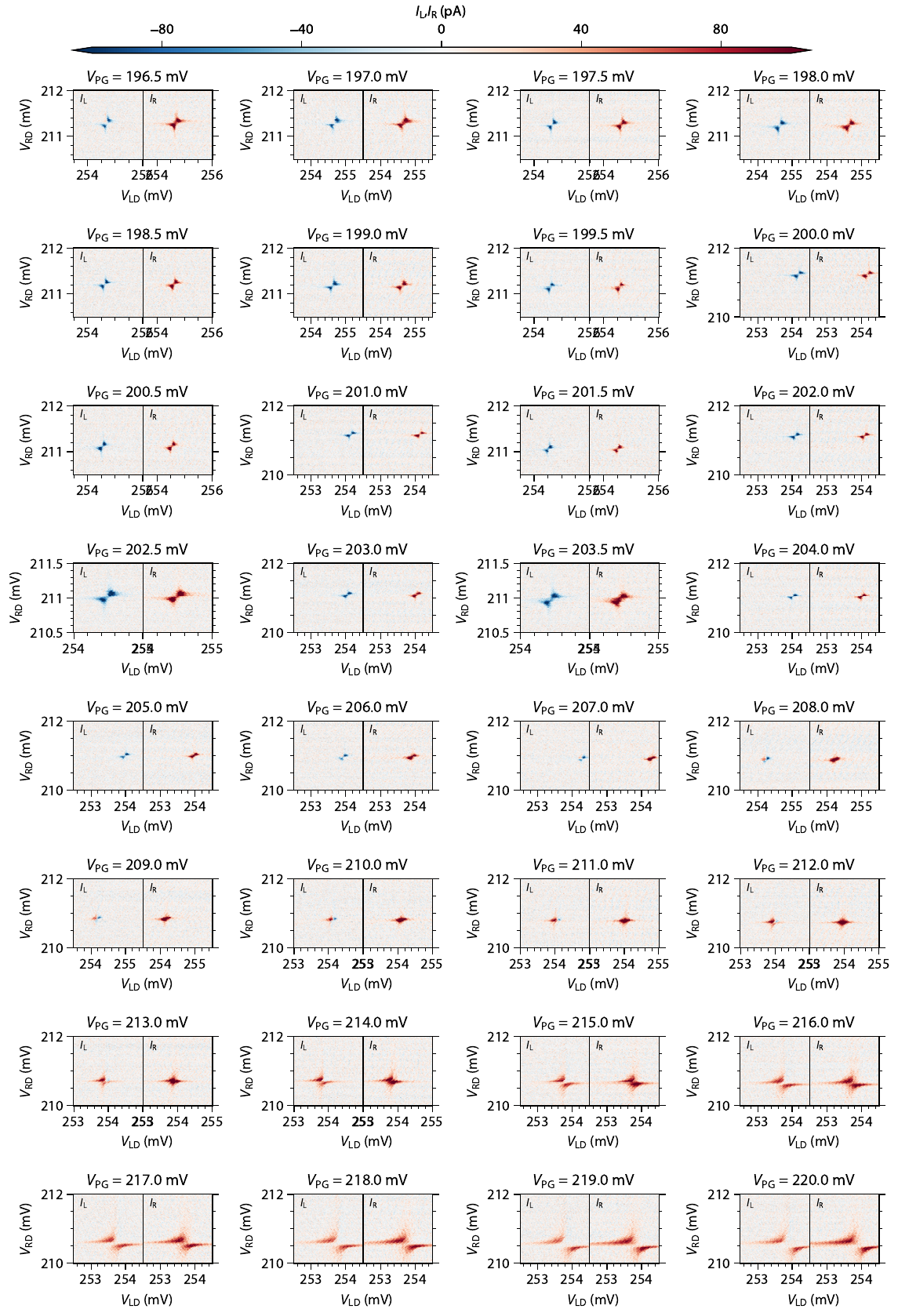}
\caption{
\textbf{Evolution of the charge stability diagram for the $\downarrow \uparrow$ spin configuration.} Each panel shows $\IL$ (nonlocal) and $\IR$ (local) as functions of $\VLD,\VRD$ measured under fixed biases $\VL=0,\VR=\SI{10}{\micro V}$. $\VPG$ is tuned from \SI{196.5}{mV}, showing signatures of the $t>\Delta$ in both local and nonlocal currents, to \SI{220}{mV}, featuring the opposite $t < \Delta$ regime. }
\label{fig:pmm-ED_VPG_DU}
\end{figure}


\begin{figure}[htbp]
\centering
\includegraphics[width=\textwidth]{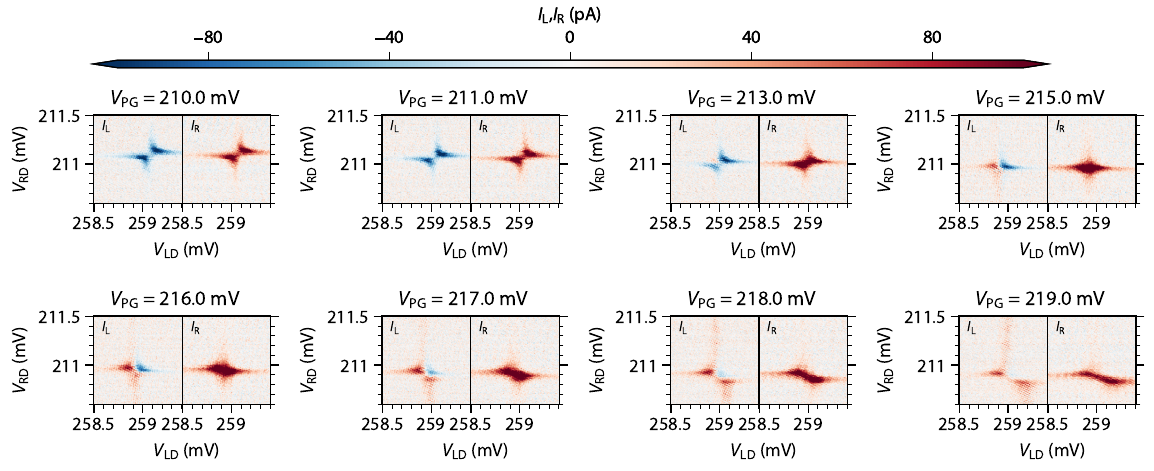}
\caption{
\textbf{Evolution of the charge stability diagram for the $\uparrow \uparrow$ spin configuration.} Each panel shows $\IL$ (nonlocal) and $\IR$ (local) as functions of $\VLD,\VRD$ measured under fixed biases $\VL=0,\VR=\SI{10}{\micro V}$. $\VPG$ is tuned from \SI{210}{mV}, showing signatures of the $t>\Delta$ in both local and nonlocal currents, to \SI{219}{mV}, featuring the opposite $t < \Delta$ regime. }
\label{fig:pmm-ED_VPG_UU}
\end{figure}


\begin{figure}[ht]
\centering
\includegraphics[width=\textwidth]{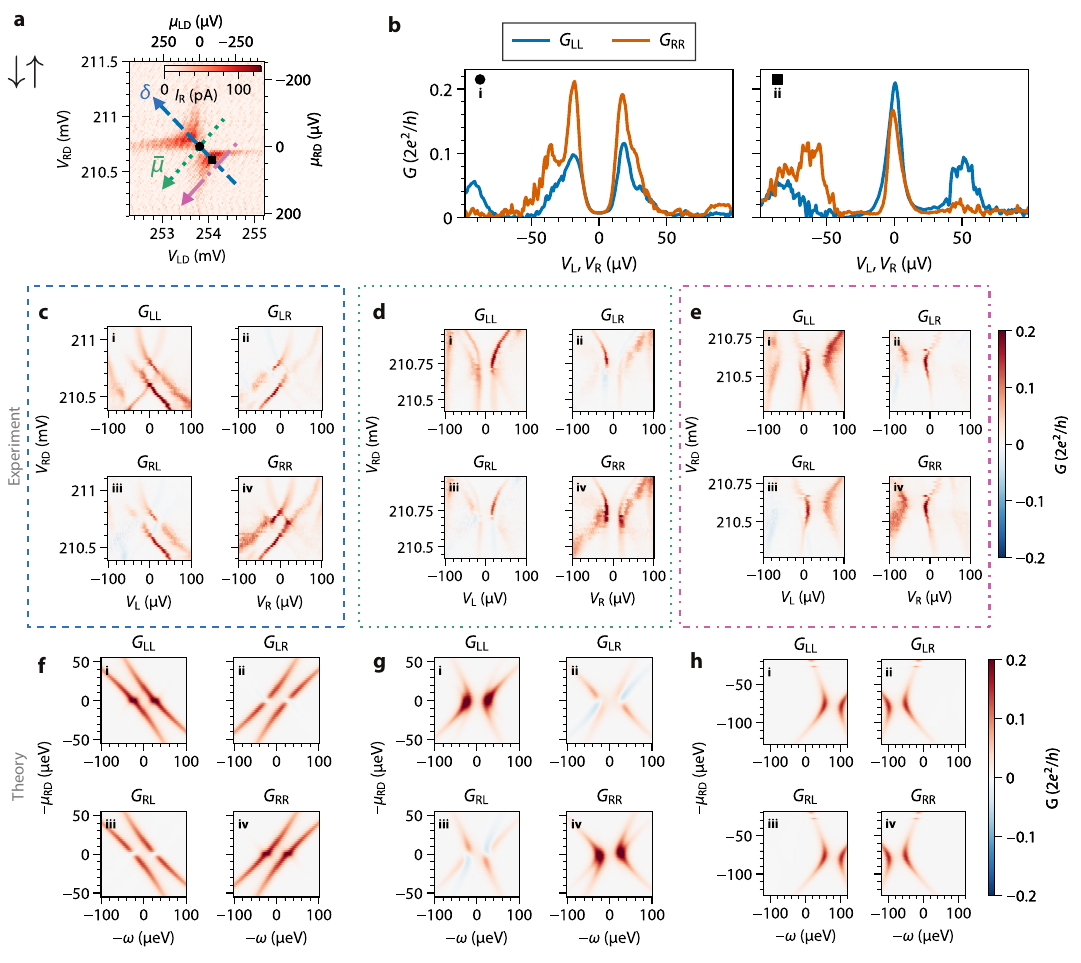}
\caption{
\textbf{Conductance spectroscopy when $t < \Delta$. } 
\textbf{a.} $\IR$ vs $\muL,\muR$ with $\VR = \SI{10}{\micro V}$. The evolution of the spectrum with the chemical potential is taken along the dashed, dashed-dotted and dotted lines in panels b,c,d respectively. Data taken at the $\downarrow \uparrow$ spin configuration with fixed $\VPG = \SI{218}{mV}$.
\textbf{b.} Local conductance spectroscopy taken at gate setpoints marked by corresponding symbols in panel~a.
Insets mark schematically the spectrum of the QDs in the absence (brown dots) and the presence (grey lines) of hybridization via CAR and ECT. 
\textbf{c.} Conductance matrix as a function of bias and $\VLD$, taken along the dashed blue line in panel~a, i.e., varying the detuning between the QDs $\delta = (\muL-\muR)/2$ while keeping the average chemical potential $\bar{\mu} = (\muL+\muR)/2$ close to 0. 
\textbf{d.} Conductance matrix as a function of bias and $\VLD$, taken along the dotted green line in panel~a, keeping the detuning between the QDs around 0. 
\textbf{e.} Conductance matrix as a function of bias and $\VLD$, taken along the dashed-dotted pink line in panel~a, keeping roughly constant non-zero detuning between the QDs.
\textbf{f, g, h.} Numerically calculated $G$ as a function of energy $\omega$ and $\muL,\muR$ along the paths shown in panel~a. 
All of the numerical curves assume the same parameters as those in \cref{fig:pmm-fig3}, except with $\Delta = \SI{23}{\micro eV}$ and $t = \SI{6}{\micro eV}$.}
\label{fig:pmm-ED_NL_ABS}
\end{figure}


\begin{figure}[ht]
\centering
\includegraphics[width=\textwidth]{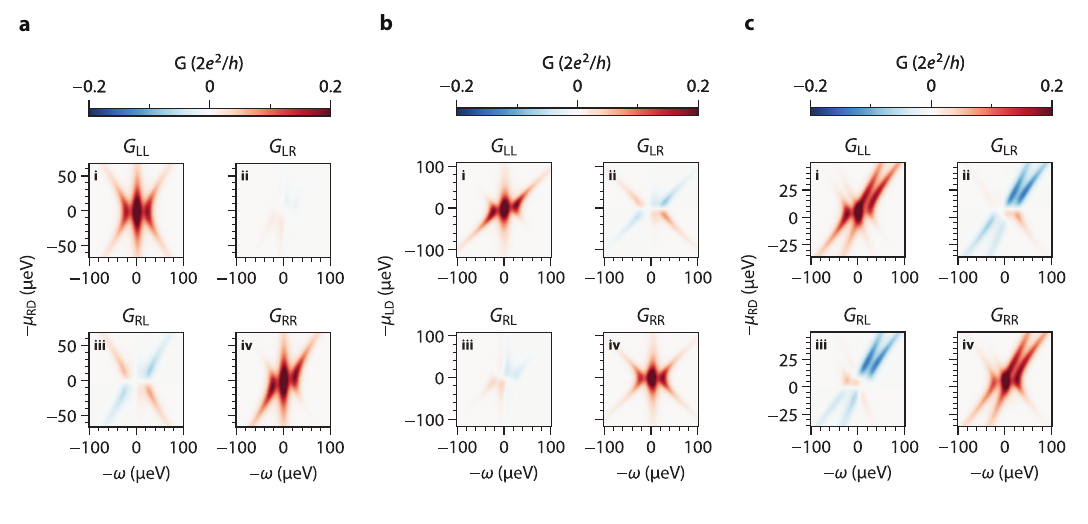}
\caption{
\textbf{Calculated conductance matrices at the $t=\Delta$ sweet spot } 
\textbf{a.} Numerically calculated $G$ as a function of energy $\omega$ and $\muL,\muR$ along the path shown in Fig~3c. The presence of finite $\GLR$ and asymmetric $\GRL$ result from a slight deviation from the $\muL=0$ condition which is depicted in \cref{fig:pmm-fig4}a. These features appear in the experimental data shown in \cref{fig:pmm-fig3}c.
\textbf{b.} Numerically calculated $G$ as a function of energy $\omega$ and $\muL,\muR$ along the path shown in Fig~3d. The presence of finite $\GRL$ and asymmetric $\GLR$ result from a slight deviation from the $\muR=0$ condition which is depicted in \cref{fig:pmm-fig4}b. These features appear in the experimental data shown in \cref{fig:pmm-fig3}d.
\textbf{c.} Numerically calculated $G$ as a function of energy $\omega$ and $\muL,\muR$ along the path shown in Fig~3e. Since the path does not obey $\muL=\muR$, the calculated spectral lines do not follow parallel trajectories, in slight disagreement with the experimental data.
The conversion from $\VLD,\VRD$ to $\muL,\muR$ is done as explained in the Methods section with the measured lever-arms of both QDs.}
\label{fig:pmm-ED_alt_fig4}
\end{figure}


\begin{figure}[htbp]
\centering
\includegraphics[width=0.8\textwidth]{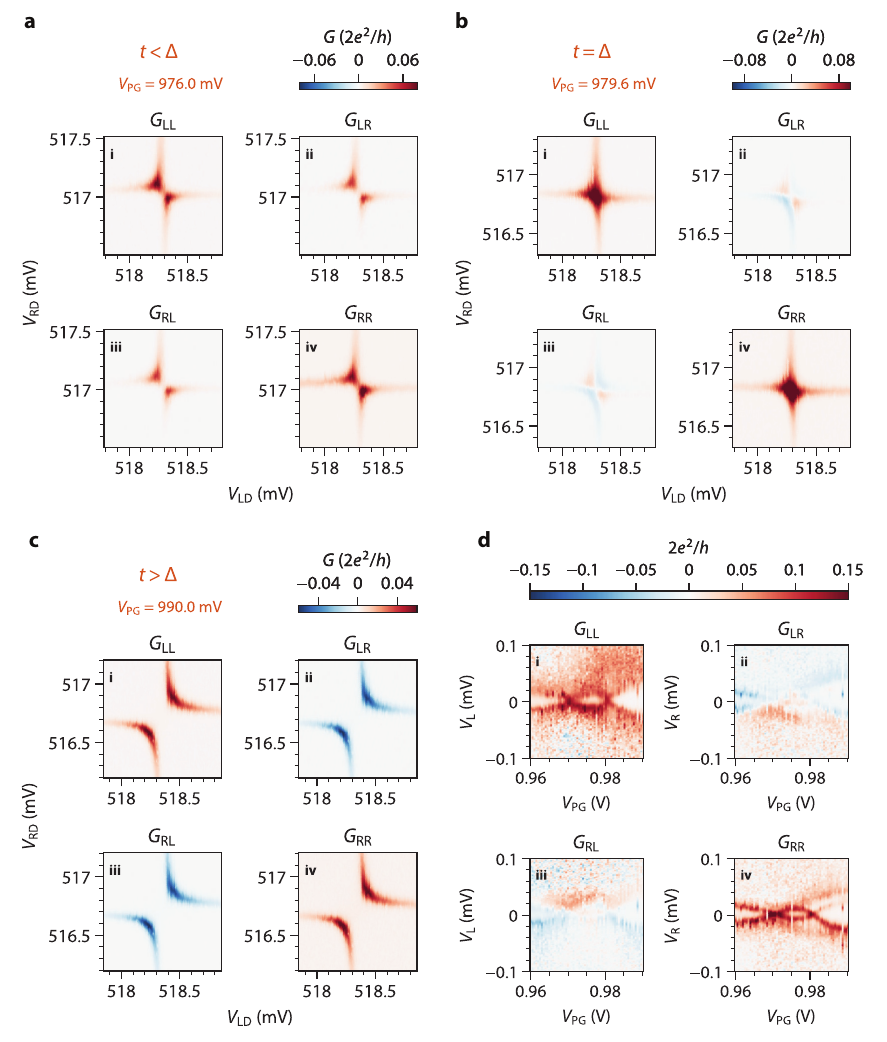}
\caption{
\textbf{Reproduction of the main results with Device~B.} \textbf{a-c.} Conductance matrices measured at  $\VPG = (976, 979.6, 990)~\si{mV}$, respectively. 
\textbf{d.} Conductance matrix as a function of $\VL,\VR$ and $\VPG$ while keeping $\muL \approx \muR \approx 0$. This device shows two continuous crossovers from $t>\Delta$ to $t<\Delta$ and again to $t>\Delta$. }
\label{fig:pmm-ED_device_2}
\end{figure}


\begin{figure}[htbp]
\centering
\includegraphics[width=\textwidth]{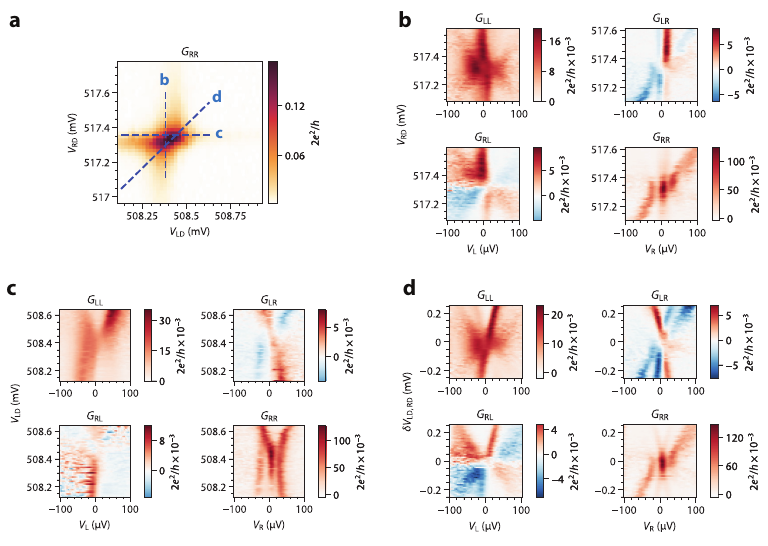}
\caption{
\textbf{Device~B spectrum vs gates.} \textbf{a.} Charge stability diagram measured via $\GRR$ of another $t=\Delta$ sweet spot of Device B, at $\VPG=\SI{993}{mV}$. Dashed lines mark the gate voltage paths the corresponding panels are taken along. 
\textbf{b-d.} Conductance matrices when varying $\VRD$ (b), $\VLD$ (c) and the two gates simultaneously (d), similar to \cref{fig:pmm-fig3} in the main text. The sticking zero-bias conductance peak feature when only one QD potential is varied around the sweet spot is clearly reproduced in $\GRR$ of panel~b. The quadratic peak splitting profile when both QD potentials are varied by the same amount is also reproduced the panel~d. The left N contact of this device was broken and a distant lead belonging to another device on the same nanowire was used instead. This and gate jumps in $\VRD$ complicate interpretation of other panels.
}
\label{fig:pmm-ED_devB_biasGates}
\end{figure}


\begin{figure}[htbp]
\centering
\includegraphics[width=\textwidth]{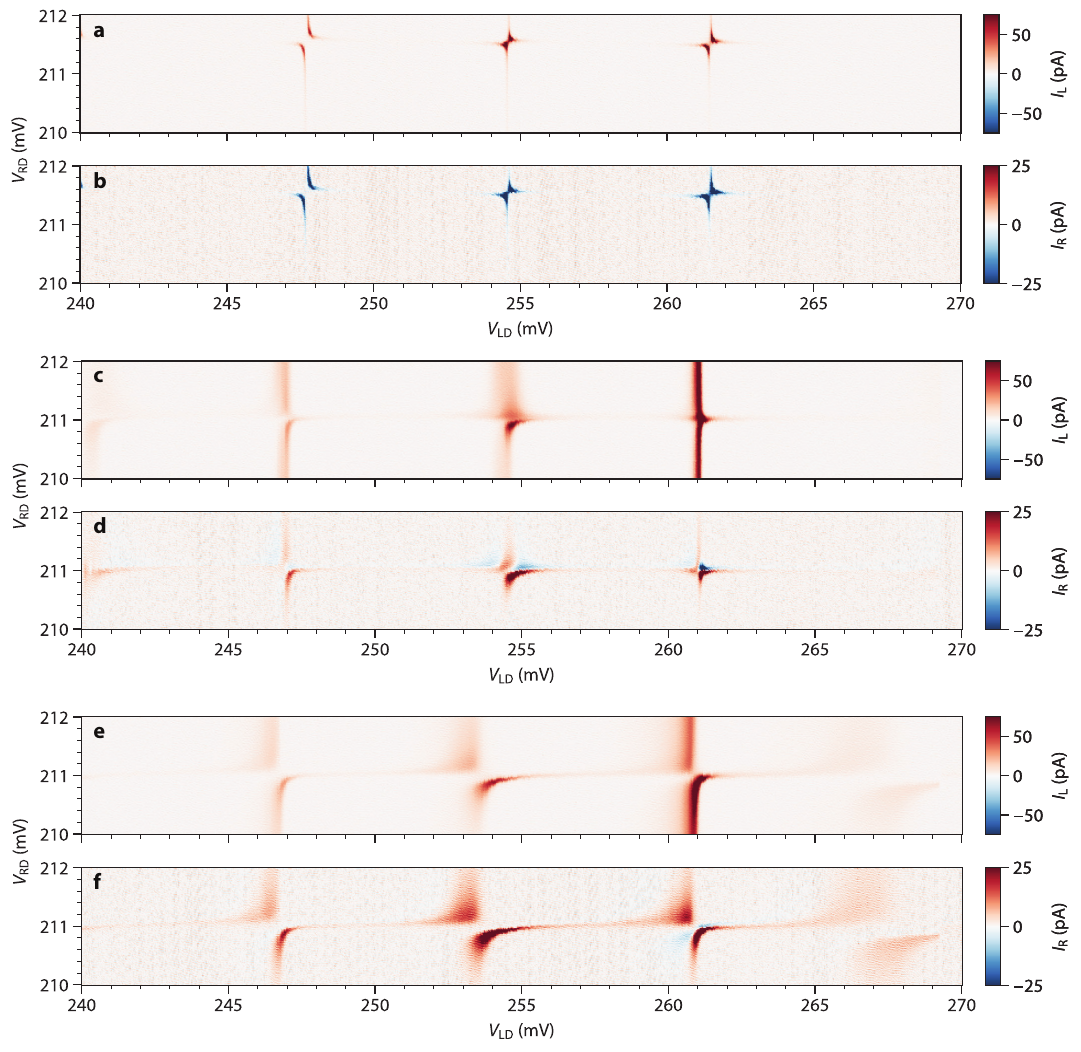}
\caption{
\textbf{CAR- and ECT-induced interactions across multiple QD resonances.} \textbf{a-b.} local ($\IL$) and nonlocal ($\IR$) currents as a function of $\VLD$ and $\VRD$ measured with $\VPG = \SI{200}{mV}$ and fixed $\VL$. All resonances show an ECT-dominated structure and a negative correlation between the local and the nonlocal currents.
\textbf{c-d.} local ($\IL$) and nonlocal ($\IR$) currents as a function of $\VLD$ and $\VRD$ measured with $\VPG = \SI{218}{mV}$ and fixed $\VL$. Some resonances show the structure associated with the $t=\Delta$ sweet spot, showing both positive and negative correlations between the local and nonlocal currents.
\textbf{e-f.} local ($\IL$) and nonlocal ($\IR$) currents as a function of $\VLD$ and $\VRD$ measured with $\VPG = \SI{200}{mV}$ and fixed $\VL$. All orbitals show a CAR-dominated structure and a positive correlation between the local and the nonlocal currents. 
All measurements were conducted with $\VL = \SI{10}{\micro V}$, $\VR = 0$ and $B = \SI{100}{mT}$.}
\label{fig:pmm-ED_multi_res}
\end{figure} 


\begin{figure}[htbp]
\centering
\includegraphics[width=\textwidth]{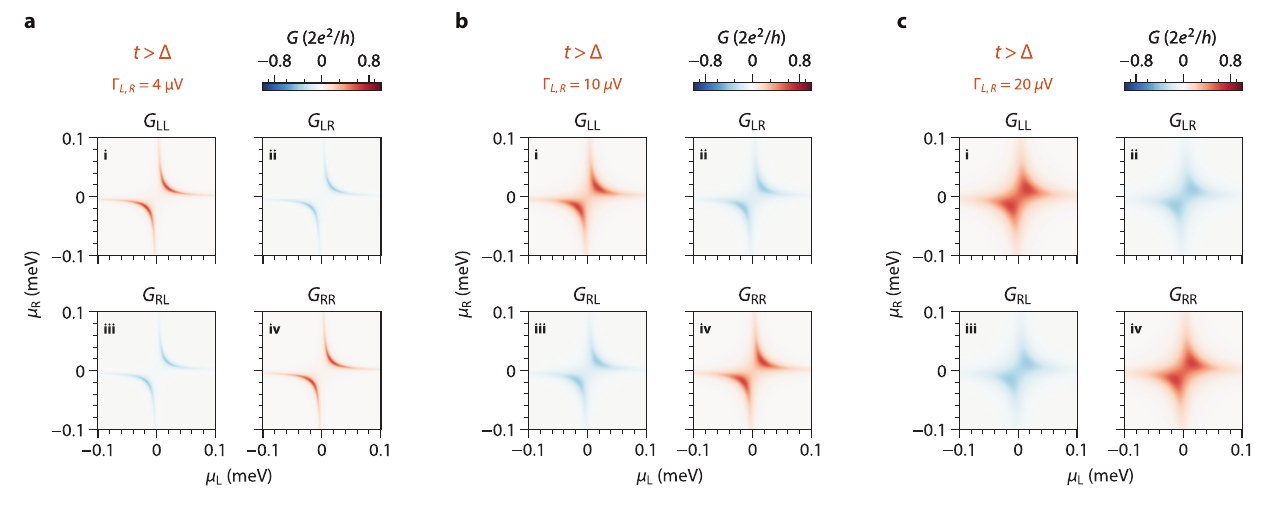}
\caption{
\textbf{Theoretical effect of tunnel broadening on the charge stability diagrams.}
In some charge stability diagrams where level-repulsion is weak, e.g., \cref{fig:pmm-fig2}a and \cref{fig:pmm-ED_VPG_UU}, some residual conductance is visible even when $\muL=\muR=0$.
This creates the visual feature of the two conductance curves appearing to ``touch'' each other at the center.
In the main text, we argued this is due to level broadening.
Here, we plot the numerically simulated charge stability diagrams at zero temperature under various dot-lead tunnel coupling strengths.
We use coupling strengths $t=\SI{20}{\micro V},\Delta=\SI{10}{\micro V}$ as an example.
From panel~a to c, increasing the tunnel coupling and thereby level broadening reproduces this observed feature. 
When the level broadening is comparable to the excitation energy, $\lvert t-\Delta \rvert$, finite conductance can take place at zero bias. 
This feature is absent in, e.g., \cref{fig:pmm-fig2}c, where $\lvert t-\Delta \rvert$ is greater than the level broadening. }
\label{fig:pmm-ED_broadening_CSD}
\end{figure}

\clearpage

\bibliography{pmm.bib}

\end{document}